\documentclass[a4paper]{article}

\usepackage[dvips]{graphicx}

\usepackage{amsmath}

\newcommand{\vc}[1]{\boldsymbol{#1}}
\def\Tr{^\mathrm{T}}
\def\proofend{\hfill\rule{8pt}{8pt}}
\def\Real{{\rm I\!R}}

\begin{document}

\title{Potential Games for Distributed Constrained Consensus}

\author{Dimitris~Ampeliotis and Kostas~Berberidis}

\date{Department of Computer Engineering and Informatics\\University of Patras, Rio - Patras, Greece\\e-mails:\{ampeliot, berberid\}@ceid.upatras.gr}

\maketitle

\begin{abstract}
The problem of computing a common point that lies in the intersection of a finite number of closed convex sets, each known to one agent in a network, is studied. This issue, known as the distributed convex feasibility problem or the distributed constrained consensus problem, constitutes an important research goal mainly due to the large number of possible applications. In this work, this issue is treated from a game theoretic viewpoint. In particular, we formulate the problem as a non-cooperative game for which a potential function exists and prove that all Nash equilibria of this game correspond to consensus states. Based upon this analysis, a best-response based distributed algorithm that solves the constrained consensus problem is developed. Furthermore, one more approach to solve the convex feasibility problem is studied based upon a projected gradient type algorithm that seeks the maximum of the considered potential function. A condition for the convergence of this scheme is derived and an exact distributed algorithm is given. Finally, simulation results for a source localization problem are given, that validate the theoretical results and demonstrate the applicability and performance of the derived algorithms.
\end{abstract}

\section{Introduction}\label{sec:intro}
Recent technological advancements, mainly in the fields of electronics and wireless communications, have enabled the development of various networks of miniaturized devices that surround us in our everyday lives. Examples of such networks include wireless sensor networks (WSNs)\cite{wsns}, multi-agent systems (MAS) \cite{Saber2007}, machine-to-machine networks (M2M) \cite{m2m} or Internet-of-Things (IoT) \cite{iot}. They serve a multitude of scopes and applications for example in smart-cities \cite{iotcities}, the industry \cite{iotindustry}, the smart energy grid \cite{m2mgrid}, and in vehicular networks \cite{iotvech}.

The expanding use of ad-hoc networks of small-scale devices along the continuous improvement of their computational and storage capabilities, drive the demand for efficient distributed algorithms that are able to confront more and more complex problems that arise in this setting \cite{Saber2007}, \cite{subgrad}, \cite{Sayed2014}. Distributed optimization based problems such as parameter estimation \cite{Plata2015}, \cite{bibC20}, decision making / detection \cite{AlSayed2018} and learning \cite{Chouvardas2011}, \cite{Towfic_pattern_rec} have been studied by many researchers and various efficient algorithms have been developed. Such algorithms alleviate the need for transmitting all the measurements obtained by the devices to a central computer for further processing and, in many cases, they enjoy no loss in performance as compared to centralized approaches. Furthermore, distributed algorithms are in general resilient to several types of device or network failures, they are scalable in the sense that minor or no modifications are required when the network is altered or expanded, and they do not suffer from the ``single point of failure'' problem, which is inherent in centralized architectures \cite{Bertsekas1989}.

The problem of convex feasibility has a central role in applied mathematics. In can be formulated in various ways, such as finding a common point of closed and convex sets, finding a common fixed-point of nonexpansive operators, finding a common minimum of convex functionals or solving a system of variational inequalities \cite{COMBETTES1996155}. Likewise, the problem has found numerous and diverse applications, in particular in solving inverse image reconstruction problems \cite{COMBETTES1996155}, \cite{censor2008iterative}, in image restoration \cite{image_restoration} but also in source localization \cite{pocs_loc_tsp} and node/network positioning problems \cite{gholami2010positioning}, \cite{gholami2011} and in distributed dictionary learning \cite{bibC18}. On the other hand, various centralized optimization algorithms have been studied for the solution of convex feasibility problems. Two very important works on the topic are \cite{gubin1967method} and \cite{browder67}. In \cite{Bauschke1996367}, the authors provide a review of various algorithms with an attempt to unify and generalize them. Notable research has also been done towards the derivation of convergence rate results for these algorithms \cite{deutsch1984rate}, \cite{deutsch2006rate}. Also, note that the well known Gerchberg-Saxton image reconstruction algorithm \cite{gs} can be seen as a POCS method where only two convex sets are employed. More recently, an algorithm of the Newton-type has been proposed for the solution of the convex feasibility problem \cite{STOSIC2016191}. In general, the most popular method for solving the convex feasibility problem consists in the successive projection onto each of the convex sets, following a fixed periodic ordering or, at least, one in which each set is utilized at least once in a fixed number of iterations (Browder’s Admissible Control law, \cite{browder67}). In the following, we will use the name \emph{Projections Onto Convex Sets} (POCS) to refer to the scheme that performs a sequence of projections while utilizing a finite number of closed convex sets in a periodic fashion.


While several centralized algorithms for the convex feasibility problem have appeared and their convergence and convergence rate behavior has been studied, literature on distributed algorithms for the convex feasibility problem is rather limited. In particular, there are only two works that develop and study such distributed algorithms, \cite{marden2007connections} and \cite{Nedic2010}. In \cite{marden2007connections}, the authors make a connection between problems that arise in cooperative control and potential games. In contrast to that work, we specialize the considered scenario by considering convex constraint sets. Furthermore, our work considers continuous sets, i.e., non-countable infinite possible strategies for each player. Also, while the work in \cite{marden2007connections} focuses on games with suboptimal Nash equilibria, in the game considered here we prove that the considered game has no Nash equilibria other than those that correspond to consensus states for the network. In \cite{Nedic2010}, the authors develop consensus and optimization algorithms for multi-agent networks. Although the work is very general, encompassing even time varying communication graphs, the derived algorithms rely upon properly selected combination weights that are not easy to select. In contrast to that work, we do not consider time-varying communication graphs, and derive algorithms with explicitly given combination weights which are easier to implement. Also, different from \cite{Nedic2010}, the algorithms proposed in this work are derived by following a game theoretic viewpoint.

The main contribution of this work is that it derives two easily implementable, yet provably convergent, algorithms for the solution of the distributed convex feasibility problem. Furthermore, this work offers a game-theoretic viewpoint to this classical mathematical problem. The proposed algorithms posses different properties. For example, the first of the proposed algorithms does not require the tuning of any additional parameters / combination weights for its operation. However, it does not allow the simultaneous strategy update of all the nodes, but rather it employs a mechanism to identify the nodes that can update their strategies at each iteration. On the other hand, the second algorithm allows all nodes to simultaneously update their strategies, however it relies upon the selection of a learning rate scalar parameter common for all the agents in the network, that must be properly adjusted. 


The remaining of this paper is organized as follows. In Section \ref{sec:pf} the problem formulation is given. In Section \ref{sec:game}, the considered problem is modelled as a non-cooperative game in strategic form and a best-response based distributed algorithm is developed. In Section \ref{sec:ga}, a gradient ascend based algorithm is proposed and a sufficient condition for its convergence is given. Finally, Section \ref{sec:nr} presents various simulation results to demonstrate the performance of the derived algorithms and Section \ref{sec:c} concludes this work.

\section{Problem Formulation}\label{sec:pf}
Consider a set of agents/nodes $\mathcal{N}=\{1,2,\ldots,N\}$ that are able to communicate as described by some time invariant, connected and undirected graph $\mathcal{G}=(\mathcal{N},\mathcal{E})$, where the set of edges $\mathcal{E}$ contains pairs of nodes that can communicate directly. Each node $n\in\mathcal{N}$ is able to communicate directly with a subset of nodes $\mathcal{N}_n \subset \mathcal{N}$, which are termed neighbours of node $n$, while $n \notin \mathcal{N}_n$. Consider also that each node $n\in\mathcal{N}$ has a closed convex set $\mathcal{C}_n \subset \Real^q$. The information that such sets represent may vary with the application considered, as an example, they could represent the node's constraints/knowledge regarding some parameter vector $\vc{p}_o \in \Real^q$ that the network is interested to estimate, in the sense that the measurements of node $n$ imply that $\vc{p}_o \in \mathcal{C}_n$. Assuming that the intersection $\mathcal{I}$ of all such sets is non-empty, i.e.,
\begin{equation}
\mathcal{I}=\bigcap_{n=1}^N\mathcal{C}_n\neq \emptyset\ ,
\end{equation}
the scope is to make all nodes agree (consent) on a common vector $\hat{\vc{p}}_o\in\mathcal{I}$, that combines the knowledge of all the agents in the network and constitutes the estimate of the network for the unknown true parameter vector $\vc{p}_o$. Furthermore, the computation of such a vector $\hat{\vc{p}}_o \in \mathcal{I}$ must be performed in a distributed fashion, in the sense that only (limited) local information exchange is possible among the nodes.

\section{Game-Theoretic Analysis}\label{sec:game}
In this section, starting from the modeling in \cite{marden2007connections}, we formulate the distributed constrained consensus problem defined in Section \ref{sec:pf} as a non-cooperative game in strategic form \cite{osborne1994course}. In the sequel, we demonstrate that an exact potential function exists for the considered game \cite{monderer1996potential}. Furthermore, we provide a proof that the considered game has no \emph{Nash equilibria} other than those that correspond to consensus states for the network. Finally, a distributed algorithm based on best-response dynamics is derived that is guaranteed to converge to such a consensus state.


\subsection{Game definition}\label{sec:game_def}
Consider a game, where the set of players is the set of nodes $\mathcal{N}$, the convex set $\mathcal{C}_n\subset \Real^q$ constitutes the so-called \emph{action set} of node $n$ and each point $\vc{p}_n \in \mathcal{C}_n$ is an action or \emph{strategy} for player $n$. It is convenient to also define the Cartesian product set $\mathcal{C}=\mathcal{C}_1\times\mathcal{C}_2\times\cdots\mathcal{C}_N$ that contains all possible selections of strategies, one for each player, i.e., all possible \emph{strategy profiles} \cite{osborne1994course}. Considering a strategy profile $\vc{p}\in\mathcal{C}$, we also denote as $\vc{p}_{-n}$ the strategies of all nodes except $n$ and write $\vc{p}=(\vc{p}_n,\vc{p}_{-n})$. In order to model the considered consensus problem as a game, each player is assigned a so-called \emph{utility function} $U_n:\mathcal{C}\rightarrow\Real$, that we define to be
\begin{equation}
U_n(\vc{p})=U_n(\vc{p}_n,\vc{p}_{-n})=-\sum_{k\in\mathcal{N}_n}\|\vc{p}_n-\vc{p}_k\|^2,\quad\vc{p}\in\mathcal{C}\ .
\label{eq:util}
\end{equation}
It is clear that the considered utility functions promote the agreement among the nodes, since a greater utility is given when neighbouring nodes select strategies close to each other. Also, it is easy to see that in the case of consensus, all nodes reach their maximum utility, which is equal to zero. Furthermore, when all nodes get zero utility then a consensus has been reached among the nodes of the network since the communication graph $\mathcal{G}$ is assumed to be connected.

\subsection{Analysis of the considered game}\label{sec:game_anal}
In this subsection we analyse the properties of the game defined in Subsection \ref{sec:game_def}. We start by characterising the utility functions in equation (\ref{eq:util}).

\noindent{\bf Lemma 1:} The utility functions 
\begin{equation}
U_n(\vc{p})=-\sum_{k\in\mathcal{N}_n}\|\vc{p}_n-\vc{p}_k\|^2,\quad\vc{p}\in\mathcal{C}\ ,
\end{equation}
are smooth and concave, or equivalently, $-U_n(\vc{p})$ are smooth, convex functions defined over the convex set $\mathcal{C}$.

\noindent{\bf Proof:} Consider the function $f_n(\vc{p})=\|\vc{p}_n\|^2$, where $\vc{p}_n$ is a sub-vector of $\vc{p}$. Clearly, $f_n(\vc{p})$ is convex in $\vc{p}_n$, since it is simply the square Euclidean norm of $\vc{p}_n$. Also, it is unaffected by the variables (elements) of $\vc{p}$ not included in $\vc{p}_n$, thus it is constant (affine) for these variables, and in total, it is convex in $\vc{p}$. Consider now the function $g_{nk}(\vc{p})=\|\vc{p}_n-\vc{p}_k\|^2$. Clearly, $g_{nk}(\vc{p})$ is convex, since it results from the composition of $f_n$ with an affine mapping \cite{hiriart2012fundamentals}. Also, $-U_n(\vc{p})$ is convex as the sum of convex functions $g_{nk}(\vc{p})=\|\vc{p}_n-\vc{p}_k\|^2$, for $k\in \mathcal{N}_n$. Furthermore, the gradient of the utility function
\begin{equation}
\nabla_n U_n(\vc{p}) = - 2\sum_{k\in\mathcal{N}_n}(\vc{p}_{n}-\vc{p}_{k})\ ,
\end{equation}
exists for all $\vc{p}\in\mathcal{C}$, thus $U_n(\vc{p})$ is a smooth function. Of course, the domain $\mathcal{C}$ is a convex set, since it is the Cartesian product of convex sets \cite{hiriart2012fundamentals}.\hfill\proofend

Consider now the function $\phi:\mathcal{C}\rightarrow\Real$, defined as
\begin{equation}
\phi(\vc{p})=-\sum_{n\in \mathcal{N}} \sum_{k \in \mathcal{N}_n} \frac{1}{2}\|\vc{p}_n-\vc{p}_k \|^2,\quad\vc{p}\in\mathcal{C}\ .
\label{eq:pf}
\end{equation}

\noindent{\bf Corollary 1:} From Lemma 1, we have that $-\phi(\vc{p})$ is a smooth, convex function defined over the convex set $\mathcal{C}$.

As shown in the following, it can be proven that the function in (\ref{eq:pf}) constitutes a so-called \emph{exact potential function} for the game defined in Subsection \ref{sec:game_def}. An exact potential function \cite{monderer1996potential} associated with a non-cooperative game, is a function with the property that whenever any single player changes their strategy and thus changes their utility function, the potential function changes by the exact same amount. Thus, games for which an upper bounded potential function exists (i.e., \emph{potential games}), have the interesting property that any sequence of selfish, greedy steps, in which all players participate while one player at a time increases their utility function, will converge to a so-called \emph{Nash equilibrium} of the game, in which no player has an incentive to change their strategy unilaterally. This point corresponds to a so-called Pareto optimal point of the potential function. In the following lemma, we provide a somewhat more general proof that (\ref{eq:pf}) is a potential function for the considered game.

\noindent{\bf Lemma 2:} Consider any fixed set of nodes $\mathcal{K}\subset\mathcal{N}$ that are not directly connected to each other in $\mathcal{G}$. Assume that all nodes in $\mathcal{K}$ update their strategies while all other nodes keep their strategies unaltered, so that the strategy profile is modified from $\vc{p}^{(1)}$ to $\vc{p}^{(2)}$. Then,
\begin{equation}
\phi(\vc{p}^{(2)})-\phi(\vc{p}^{(1)})=\sum_{n\in\mathcal{K}}\left(U_n(\vc{p}^{(2)})-U_n(\vc{p}^{(1)})\right)\ .
\end{equation}

\noindent{\bf Proof:} We can write the function in (\ref{eq:pf}) using two terms, where the first one depends on the strategies of the nodes $n\in\mathcal{K}$ while the second term does not depend on these strategies, as
\begin{eqnarray}
&&\phi(\vc{p})=\nonumber\\
&=&-\sum_{n\in \mathcal{K}} \sum_{k \in \mathcal{N}_n} \|\vc{p}_n-\vc{p}_k \|^2
-\sum_{n\in \mathcal{N}\setminus\mathcal{K}} \sum_{k \in \mathcal{N}_n\setminus\mathcal{K}} \frac{\|\vc{p}_n-\vc{p}_k \|^2}{2}\nonumber\\
&=&\sum_{n\in \mathcal{K}}U_n(\vc{p})-\sum_{n\in \mathcal{N}\setminus\mathcal{K}} \sum_{k \in \mathcal{N}_n\setminus\mathcal{K}} \frac{\|\vc{p}_n-\vc{p}_k \|^2}{2}
\label{eq:pfsep}
\end{eqnarray}
The second term in (\ref{eq:pfsep}) does not depend on the strategies that change, thus, it has the same value for $\vc{p}^{(1)}$ and $\vc{p}^{(2)}$. Based upon this observation, we can write that
\begin{equation}
\phi(\vc{p}^{(2)})-\phi(\vc{p}^{(1)}) = \sum_{n\in\mathcal{K}}\left(U_n(\vc{p}^{(2)})-U_n(\vc{p}^{(1)})\right)\ ,
\end{equation} 
which concludes our proof.\hfill\proofend

\noindent{\bf Corollary 2:} Using Lemma 2 it is easy to note that, as a special case, when $\mathcal{K}=\{n\}$, we have the proof that $\phi(\vc{p})$ is an exact potential function for the considered game.

\noindent{\bf Theorem 1:} The game defined in Subsection \ref{sec:game_def} has no Nash equilibria other than those that correspond to consensus states.

\noindent{\bf Proof:}
Consider that the strategy profile $\vc{p}_a$ is a Nash equilibrium that does not correspond to consensus, thus
\begin{equation}
\phi(\vc{p}_a)<0\ .
\end{equation}
Denoting as $\vc{p}_{a,n}$ the strategy of node $n$ in $\vc{p}_a$ and given that $\vc{p}_a$ is a Nash equilibrium, then, no player is able to increase their utility by changing their strategy unilaterally. Since the utility functions are smooth, consider that node $n$ employs the constant step size version of the gradient projection algorithm \cite{Bertsekas1989} to increase their utility. We have assumed that a Nash equilibrium has been reached, thus the gradient projection algorithm at each node $n$ should have reached a limit point in the sense that for some suitable\footnote{It is easy to verify that any $s$ given by Lemma 4 can be considered.} $s>0$ we would have 
\begin{equation}
\vc{p}_{a,n}=\mathcal{P}_{\mathcal{C}_n}\left(\vc{p}_{a,n}+s \nabla_n U_n(\vc{p}_a) \right)\ ,
\label{eq:gp1}
\end{equation}
where $P_{\mathcal{C}_n}(\cdot)$ denotes the projection operator onto the set $\mathcal{C}_n$ and $\nabla_n$ denotes the gradient vector at the entries of $\vc{p}_{a,n}$. Following the definitions for the utility and potential functions, it can be verified that
\begin{equation}
\nabla_n U_n(\vc{p}_a) = \nabla_n \phi(\vc{p}_a)=-2\sum_{k\in\mathcal{N}_n}(\vc{p}_{a,n}-\vc{p}_{a,k})\ .
\end{equation}
Thus, equation (\ref{eq:gp1}) can be written as
\begin{equation}
\vc{p}_{a,n}=\mathcal{P}_{\mathcal{C}_n}\left(\vc{p}_{a,n}+s \nabla_n \phi(\vc{p}_a) \right)\ .
\label{eq:gp2}
\end{equation}
Finally, since projecting onto each of the convex sets $\mathcal{C}_n$ is equivalent to the projection onto the Cartesian product set $\mathcal{C}$ (See \cite{dattorro2010convex}, Example E.10.2.0.3), the equations in (\ref{eq:gp2}) for each $n$ can be summarized into one, independent of $n$, as
\begin{equation}
\vc{p}_{a}=\mathcal{P}_{\mathcal{C}}\left(\vc{p}_{a}+s \nabla \phi(\vc{p}_a) \right)\ .
\label{eq:gp3}
\end{equation}
Equation (\ref{eq:gp3}) can be recognised as the constant step size version of the gradient projection algorithm for the maximization of the potential function, 
and equation (\ref{eq:gp3}) suggests also that $\vc{p}_{a}$ is a limit point. Thus, it implies that $\vc{p}_a$ is a local maximum of the potential function $\phi(\vc{p})$. However, since the potential function is concave and its maximum is equal to zero (at the consensus points), it cannot have any local maxima with $\phi(\vc{p}_a)<0$. Thus, we arrive to a contradiction, meaning that the considered game cannot have any Nash equilibria other than those that correspond to consensus states.
\hfill\proofend

\noindent{\bf Corollary 3:} From Theorem 1, it is easy to verify that any (distributed) algorithm that computes Nash equilibria for the game defined in Subsection \ref{sec:game_def} will converge to a consensus state, thus solving the (distributed) constrained consensus problem.

In the following, we propose a distributed algorithm for computing Nash equilibria for the game defined in Subsection \ref{sec:game_def} that is based upon best-response dynamics. Furthermore, the algorithm takes advantage of Lemma 2 to allow the simultaneous update of the strategies of nodes that are not directly connected to each other in $\mathcal{G}$. As it will be seen, only local messages are required by the proposed algorithm for reaching a consensus state. 


\subsection{Algorithm development}
Let us consider the sub-problem of updating the strategy of one node, say $n$, given that the other nodes of the network (at least, the neighbours of node $n$) do not alter their strategies. We are interested in computing the best possible strategy for node $n$, defined as the one that maximizes their utility function, an approach known as \emph{best response}. Computing the best possible strategy for player $n$ given that the other players of the network have selected their strategies according to the strategy profile $\vc{p}^{(t)}$ at some discrete time instant $t$, amounts to solving the following optimization problem
\begin{eqnarray}
\vc{p}_n^{(t+1)}&=&
\arg \max_{\vc{p}_n\in\mathcal{C}_n}\left(U_n(\vc{p}_n,\vc{p}_{-n}^{(t)})\right)\nonumber\\
&=&
\arg \max_{\vc{p}_n\in\mathcal{C}_n}\left(-\sum_{k\in\mathcal{N}_n}\|\vc{p}_n-\vc{p}_k^{(t)}\|^2\right)\ .
\label{eq:opt}
\end{eqnarray}
This optimization problem employs a smooth, concave cost function and a convex constraint. Thus, some suitable convex optimization method could be used for its solution. However, in the following we directly give the optimal solution.

\noindent{\bf Lemma 3:} The optimal solution of problem (\ref{eq:opt}) is given by
\begin{equation}
\vc{p}_n^{(t+1)}=\mathcal{P}_{\mathcal{C}_n}\left(\frac{1}{|\mathcal{N}_n|}\sum_{k\in\mathcal{N}_n}\vc{p}_k^{(t)}\right)\ .
\label{eq:best_resp}
\end{equation}

\noindent{\bf Proof:} We first consider the unconstrained problem. The optimal solution for the unconstrained problem is a point in $\Real^q$ that minimizes the sum of square distances from a set of given points $\vc{p}_k^{(t)}$, $k\in\mathcal{N}_n$. It can be easily verified (by setting the gradient of the function being optimized equal to zero) that the solution to this, unconstrained, problem is given simply as the \emph{centroid} of these points
\begin{equation}
\mathbf{c}_n^{(t)}=\frac{1}{|\mathcal{N}_n|}\sum_{k\in\mathcal{N}_n}\vc{p}_k^{(t)}\ .
\end{equation}
We now turn our attention to the constraint in (\ref{eq:opt}). After some algebra, we have that
\begin{equation}
U_n(\vc{p}_n,\vc{p}_{-n}^{(t)})=-|\mathcal{N}_n|\left\|\vc{p}_n-\mathbf{c}_n^{(t)}\right\|^2+C
\label{eq:level}
\end{equation}
where
\begin{equation}
C=\left|\mathcal{N}_n\right|(\mathbf{c}_n^{(t)})\Tr\mathbf{c}_n^{(t)}-\sum_{k\in\mathcal{N}_n}(\vc{p}_k^{(t)})\Tr\vc{p}_k^{(t)}
\end{equation}
is constant. Thus, we can see from (\ref{eq:level}) that the utility function changes its value according to $\|\vc{p}_n-\mathbf{c}_n^{(t)}\|^2$, which means that it is symmetric around $\mathbf{c}_n^{(t)}$, that is, the utility function has a constant value at all points with equal distance from $\mathbf{c}_n^{(t)}$. In other words, the level-sets of the utility function are hyper-spheres centered at $\mathbf{c}_n^{(t)}$. This property suggests that the point in the set $\mathcal{C}_n$ that is the closest one to the centroid $\mathbf{c}_n^{(t)}$ will be the optimal solution for the constrained problem in (\ref{eq:opt}). This closest point is given as the projection of $\mathbf{c}_n^{(t)}$ onto the convex set $\mathcal{C}_n$.\hfill\proofend

In light of Lemma 3, a possible algorithm for the solution of the (distributed) constrained consensus problem would be to utilize equation (\ref{eq:best_resp}) and follow a fixed periodic ordering for the nodes or, at least, one in which each node updates their strategy at least once in a fixed number of iterations as in \cite{browder67}. Clearly, such an approach requires some prior organization of the network and hinders the fully distributed implementation of the algorithm. Furthermore, by following such an approach only one node would update their strategy at each iteration and could lead to very slow convergence especially for large-scale networks.

To alleviate these concerns, we can exploit Lemma 2 to allow for the simultaneous updates of strategies of nodes that are not directly connected in $\mathcal{G}$. In Table \ref{alg1} we present the proposed distributed algorithm followed by each node $n$ for the solution of the distributed constrained consensus problem.  The algorithm utilizes a mechanism to select the nodes that update their strategies at each discrete time instant $t$. In particular, all nodes compute the length of the update that they can obtain
\begin{equation}
m_n^{(t)}=\left\|\vc{r}_n^{(t)}-\vc{p}_n^{(t-1)}\right\|^2\ ,
\end{equation}
where $\vc{r}_n^{(t)}$ denotes the updated strategy of node $n$ according to (\ref{eq:best_resp}), send this metric to all their neighbours and follow the (greedy) rule that only the node with the greater value for this metric in their neighbourhood will actually perform the update. In the case of a tie, the node with the highest id wins. We refer to this algorithm as the Distributed Game Theoretic Consensus (DGTC) Algorithm.

\begin{table}[t]
\begin{center}
\begin{tabular}{|l|}
\hline
Input: Convex set $\mathcal{C}_n$, Set of Neighbours $\mathcal{N}_n$, Iterations $T$\\
Output: Final point $\vc{p}_n^{(T)}$\rule{150pt}{0pt}\\
\hline
\hline
1. Compute an initial point $\vc{p}_n^{(0)}\in\mathcal{C}_n$\\
2. for $t$ = $1$ to $T$\\
3. \qquad Send $\vc{p}_n^{(t-1)}$ to neighbouring nodes in $\mathcal{N}_n$\\
4. \qquad Receive $\vc{p}_k^{(t-1)}$ from neighbours $k\in\mathcal{N}_n$\\
5. \qquad Compute\\
\qquad \qquad \qquad $\vc{r}_n^{(t)}=\mathcal{P}_{\mathcal{C}_n}\left(\frac{1}{|\mathcal{N}_n|}\sum_{k\in\mathcal{N}_n}\vc{p}_k^{(t)}\right)$\\
6. \qquad Compute the metric\\
\qquad \qquad \qquad $m_n^{(t)}=\left\|\vc{r}_n^{(t)}-\vc{p}_n^{(t-1)}\right\|^2 $\\
7. \qquad Send $m_n^{(t)}$ to neighbouring nodes in $\mathcal{N}_n$\\
8. \qquad Receive $m_k^{(t)}$ from neighbours $k\in\mathcal{N}_n$\\
9. \qquad if $m_n^{(t)}>\max_k(m_k^{(t)})$\\
10. \qquad \qquad $\vc{p}_n^{(t)}=\vc{r}_n^{(t)}$\\
11. \qquad elseif $m_n^{(t)}=\max_k(m_k^{(t)})$ and $n>\arg\max_{k\in\mathcal{N}_n}(m_k^{(t)})$\\
12. \qquad \qquad $\vc{p}_n^{(t)}=\vc{r}_n^{(t)}$\\
12. \qquad else\\
13. \qquad \qquad $\vc{p}_n^{(t)}=\vc{p}_n^{(t-1)}$\\
14. \qquad end if\\
15. end for\\
\hline
\end{tabular}
\end{center}
\caption{Distributed Game Theoretic Consensus Algorithm at node $n$}
\label{alg1}
\end{table}

\section{Gradient Ascend on the Potential Function}\label{sec:ga}

In the previous section we developed a distributed algorithm for the constrained consensus problem by following a game-theoretic optimization methodology. As we saw, that method required some mechanism to ensure that neighboring nodes do not update their strategies simultaneously. Although the developed mechanism required only minimal and local interactions, in this section we study another approach that fully alleviates the need for such a mechanism. In particular, we propose a gradient ascend algorithm that seeks the global maximum of the potential function where all nodes can update their estimates at the same time. Interestingly, the studied approach results into an algorithm very similar to the one studied in \cite{Nedic2010}. Furthermore, we provide a novel condition for the convergence of this approach when using a constant step-size. In the following, we first derive the algorithm in a centralized setting and then we show that it can be implemented, exactly, in a distributed scenario.

\subsection{Centralized approach}

In this subsection we derive an algorithm for seeking the global maximum of the potential function in (\ref{eq:pf}). Since the maximum of the potential function corresponds to consensus states for the network, its maximization will mean the solution of the constrained consensus problem. The considered problem is equivalent to the minimization of the cost function
\begin{equation}
J(\vc{p})=-\phi(\vc{p})=\sum_{n\in \mathcal{N}} \sum_{k\in \mathcal{N}_n}\frac{1}{2}\left\|\vc{p}_n-\vc{p}_k\right\|^2
\ ,\ \vc{p}\in\mathcal{C}\ .
\label{eq:cgd}
\end{equation}
According to Corollary 1, $J(\vc{p})$ is a smooth, convex function defined over the convex set $\mathcal{C}$. Thus, the gradient projection algorithm \cite{Bertsekas1989} can be used to optimize (\ref{eq:cgd}) and compute its global minimum. Employing the constant step size version of the gradient projection algorithm yields the update equation
\begin{equation}
\vc{p}^{(t+1)}=\mathcal{P}_{\mathcal{C}}\left(\vc{p}^{(t)}-s\cdot\nabla J(\vc{p}^{(t)})\right)\ ,
\label{eq:gpa}
\end{equation}
where $\vc{p}^{(t)}$ denotes the vector of all node strategies at iteration $t$ and $s$ is the constant step size. The gradient of (\ref{eq:cgd}) is given by
\begin{equation}
\nabla J(\vc{p}^{(t)})=\left[\nabla_1 J(\vc{p}^{(t)})\Tr \cdots \nabla_N J(\vc{p}^{(t)})\Tr\right]\Tr\ ,
\end{equation}
where
\begin{equation}
\nabla_n J(\vc{p}^{(t)}) = 2 \sum_{k\in\mathcal{N}_n}(\vc{p}_n^{(t)}-\vc{p}_k^{(t)})\ .
\end{equation}
Finally, the required step size $s$ can be selected via the following lemma.

\noindent{\bf Lemma 4:} Every limit point of $\{\vc{p}^{(t)}\}$ generated by the gradient projection algorithm in (\ref{eq:gpa}) with
\begin{equation}
0<s<\frac{1}{2\sqrt{q\sum_{n=1}^N \left|\mathcal{N}_n\right|^2}}
\end{equation}
is stationary.

\noindent{\bf Proof:} Consider any two strategy profiles $\mathbf{x}\in \Real^{N\cdot q}$ and $\mathbf{y}\in \Real^{N\cdot q}$ and denote their elements using two indexes, $i$ and $j$, where $\mathbf{x}_{ij}$ denotes the $j$-th element of the strategy of agent $i$, where $i=1,2,\ldots,N$ and $j=1,2,\ldots,q$, and similarly for $\mathbf{y}_{ij}$. Also, define
\begin{equation}
\delta_{ij}=\mathbf{x}_{ij}-\mathbf{y}_{ij}\ .
\end{equation}
Using these definitions, we have that
\begin{equation}
\left\|\mathbf{x}-\mathbf{y}\right\|^2 = \sum_{i=1}^N \sum_{j=1}^q \delta_{ij}^2\geq \max_{l,m}(|\delta_{lm}|^2)\ .
\end{equation}
On the other hand, we have that
\begin{eqnarray}
\left\|\nabla J(\mathbf{x})\right.-\left.\nabla J(\mathbf{y})\right\|^2&=&\sum_{i=1}^N \sum_{j=1}^q 4 \left(\sum_{k\in\mathcal{N}_i} (\delta_{ij}-\delta_{kj})\right)^2\nonumber\\
&\leq & \sum_{i=1}^N \sum_{j=1}^q 4 \left(\sum_{k\in\mathcal{N}_i} (|\delta_{ij}|+|\delta_{kj}|)\right)^2\nonumber\\
&\leq & \sum_{i=1}^N \sum_{j=1}^q 4 \left(\sum_{k\in\mathcal{N}_i} 2 \max_{l,m}(|\delta_{lm}|)\right)^2\nonumber\\
& = & \sum_{i=1}^N \sum_{j=1}^q 16 |\mathcal{N}_i|^2 \max_{l,m}(|\delta_{lm}|^2)\nonumber\\
& = & \sum_{i=1}^N 16 q |\mathcal{N}_i|^2 \max_{l,m}(|\delta_{lm}|^2)\nonumber\\
&\leq & \sum_{i=1}^N 16 q |\mathcal{N}_i|^2 \left\|\mathbf{x}-\mathbf{y}\right\|^2
\end{eqnarray}
Thus, we have that
\begin{equation}
\left\|\nabla J(\mathbf{x})-\nabla J(\mathbf{y})\right\|\leq L \left\|\mathbf{x}-\mathbf{y}\right\|\ ,
\end{equation}
with
\begin{equation}
L = 4\sqrt{q \sum_{i=1}^N |\mathcal{N}_i|^2}\ .
\end{equation}
Finally, using Proposition 2.3.2 from \cite{Bertsekas1989}, we have that every limit point of $\{\vc{p}^{(t)}\}$ generated by (\ref{eq:gpa}) using
\begin{equation}
0<s<\frac{2}{L}=\frac{1}{2\sqrt{q \sum_{i=1}^N |\mathcal{N}_i|^2}}\ ,
\end{equation}
is stationary.\hfill\proofend

\subsection{Distributed approach}

The gradient projection algorithm, described by equation (\ref{eq:gpa}) in the previous, was developed without considering distributed implementation. However, it is quite clear to see that an exact distributed implementation is possible. First, if we define for simplicity of notation
\begin{equation}
\vc{u}_n^{(t)}=\vc{p}_n^{(t)}-2s\sum_{k\in \mathcal{N}_n}(\vc{p}_n^{(t)}-\vc{p}_k^{(t)})\ ,
\label{eq:d1}
\end{equation}
then, the vector in the projection operator in (\ref{eq:gpa}) will be given by
\begin{equation}
\vc{p}^{(t)}-s\cdot\nabla J(\vc{p}^{(t)})=\left[
\begin{array}{c}
\vc{u}_1^{(t)}\\
\vdots\\
\vc{u}_N^{(t)}\\
\end{array}
\right]\ ,
\label{eq:d2}
\end{equation}
where each part $\vc{u}_n^{(t)}$ can be computed locally at node $n$ if neighbouring nodes send their strategies $\vc{p}_k^{(t)}, k\in \mathcal{N}_n$ to node $n$. Secondly, the projection operator onto the Cartesian product set $\mathcal{C}$ is equivalent to projections onto each of the constituent convex sets $\mathcal{C}_n$ (See \cite{dattorro2010convex}, Example E.10.2.0.3), as shown by
\begin{equation}
\mathcal{P}_{\mathcal{C}}\left(
\left[
\begin{array}{c}
\vc{u}_1^{(t)}\\
\vdots\\
\vc{u}_N^{(t)}
\end{array}
\right]
\right)=
\left[
\begin{array}{c}
\mathcal{P}_{\mathcal{C}_1}(\vc{u}_1^{(t)})\\
\vdots\\
\mathcal{P}_{\mathcal{C}_N}(\vc{u}_N^{(t)})\\
\end{array}
\right]\ .
\label{eq:d3}
\end{equation}

Equations (\ref{eq:d1}), (\ref{eq:d2}) and (\ref{eq:d3}) demonstrate that, in fact, the algorithm of equation (\ref{eq:gpa}) can be exactly implemented by the individual nodes of the network. Thus, given that the considered cost function is smooth and convex, the centralized gradient projection algorithm is known to converge to a global minimum point of the cost function, and since the distributed implementation is identical to the centralized one, it will have the same convergence properties. The overall distributed approach, referred to as the distributed gradient projection consensus algorithm, is summarized in Table \ref{alg2}.

\begin{table}
\begin{center}
\begin{tabular}{|l|}
\hline
Input: Convex set $\mathcal{C}_n$, Set of Neighbours $\mathcal{N}_n$, Step size $s$, Iterations $T$\\
Output: Final point $\vc{p}_n^{(T)}$\rule{150pt}{0pt}\\
\hline
\hline
1. Compute an initial point $\vc{p}_n^{(0)}\in\mathcal{C}_n$\\
2. For $t$ = $1$ to $T$\\
3. \qquad Send $\vc{p}_n^{(t-1)}$ to neighbouring nodes in $\mathcal{N}_n$\\
4. \qquad Receive $\vc{p}_k^{(t-1)}$ from neighbours $k\in\mathcal{N}_n$\\
5. \qquad Update \\
6. \qquad \quad $\vc{p}_n^{(t)}=\mathcal{P}_{\mathcal{C}_n}\left(\vc{p}_n^{(t-1)} - 2 s \sum_{k \in \mathcal{N}_n}\left(\vc{p}_n^{(t-1)}-\vc{p}_k^{(t-1)}\right)\right)$\\
7. End for\\
\hline
\end{tabular}
\end{center}
\caption{Distributed gradient projection consensus algorithm at node $n$}
\label{alg2}
\end{table}

\section{Numerical results}\label{sec:nr}
In order to validate the theoretical results derived in the previous and to study the relative convergence rate of the considered algorithms under various scenarios, some simulation results were conducted.

\subsection{Validation of convergence results}
In the first considered simulation our scope is to validate the convergence of the consensus algorithms. To this end, a network of $N=100$ nodes is simulated where each node is uniformly placed in the $q$-dimensional unit box at locations $\vc{l}_n, n=1,\ldots,N$ and direct communication is possible when the distance between two nodes is less or equal to a parameter $\rho$. The nodes of the network cooperate in a source localization problem. In particular, a source is uniformly placed at location $\vc{z}$ in the central $q$-dimensional box with side length equal to 0.5. The convex set of node $n$ is defined as the hyper-sphere centered at the location of node $n$ with radius equal to the node-source distance plus some small positive constant, i.e.,
\begin{equation}
\mathcal{C}_n = \left\{\vc{l}\in\Real^q: \left\|\vc{l}_n-\vc{l} \right\| \leq \left\|\vc{l}_n-\vc{z} \right\|+\epsilon\right\}\ ,
\end{equation}
where $\epsilon=0.01$ is used to ensure that the source is located inside $\mathcal{C}_n$. For the gradient projection consensus algorithm we use a step size close to the maximum allowed according to Lemma 4, and in particular we set
\begin{equation}
s=0.99 \frac{1}{2\sqrt{q \sum_{i=1}^N |\mathcal{N}_i|^2}}\ .
\label{sim_step}
\end{equation}
Also, we perform Monte-Carlo simulation of 50 different network instances and keep only instances of the localization problem that resulted into a connected network topology $\mathcal{G}$. For comparison, the centralized POCS scheme is included in the results where we perform 40 circles of projections with a fixed node ordering. 

In the following we consider a ``consensus metric'' $c^{(t)}$ that measures how close to each other the estimates of the nodes are and we define it as 
\begin{equation}
c^{(t)}=\sqrt{\sum_{n=1}^{N}\sum_{j=1}^{q} \left|p_{n,j}^{(t)}-\mu_j^{(t)}\right|^2}\ ,
\end{equation}
where
\begin{equation}
\mu_j^{(t)}=\frac{1}{N}\sum_{n=1}^{N}p_{n,j}^{(t)}\ ,
\end{equation}
and $p_{n,j}^{(t)}$ denotes the $j$-th component of the estimate at node $n$ at time $t$.

\begin{figure}
\begin{center}
\includegraphics[width=0.8\textwidth]{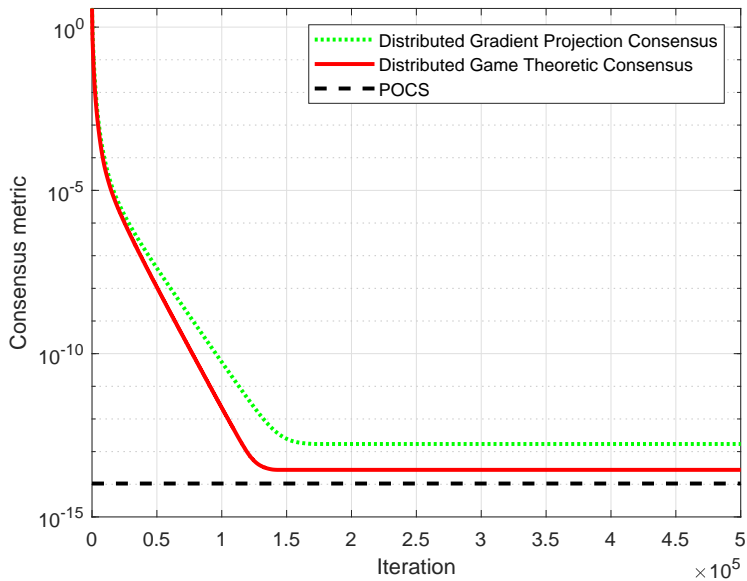}
\end{center}
\caption{Consensus metric as a function of iterations in $q=2$ dimensions, communication range $\rho=0.3$}
\label{fig1}
\end{figure}

\begin{figure}
\begin{center}
\includegraphics[width=0.8\textwidth]{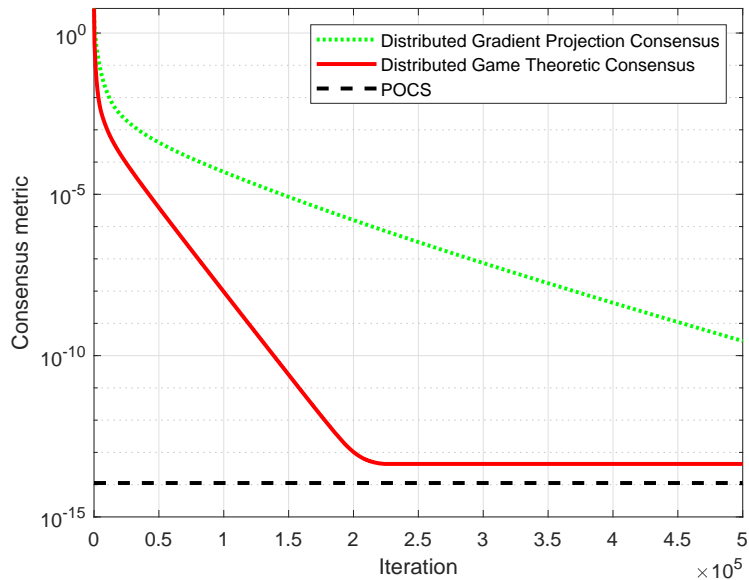}
\end{center}
\caption{Consensus metric as a function of iterations in $q=3$ dimensions, communication range $\rho=0.3$}
\label{fig2}
\end{figure}

Figures \ref{fig1} and \ref{fig2} demonstrate the evolution of the consensus metric as a function of iterations for $q=2$ and $q=3$, respectively. In both cases the communication range was $\rho=0.3$. It is evident from these plots that consensus is always reached, and the consensus metric reaches a value close to machine precision. The difference in steady-state values is due to the different numerical properties of the considered algorithms. Also, it appears that the the developed game theoretic consensus algorithm converges faster than the considered gradient projection algorithm, in these scenarios. In order to better study the convergence rate behavior of the considered algorithms further numerical results are presented in the sequel.

\subsection{Convergence rate comparisons}
In this section, the convergence rate of the considered algorithms is evaluated as a function of the network edge density. In more detail, the localization problem considered in the previous is repeated here, but 1000 network realizations are simulated that each corresponds to a different network topology and edge density. The edge density of the network is varied by using a different communication range in each realization. We measure the edge density of the network in terms of the so-called Fiedler eigenvalue \cite{brouwer2011spectra}, where smaller values correspond to more sparsely connected networks. For evaluating the rate of convergence, the number of iterations required so that the consensus metric drops below $10^{-5}$ is measured, for each algorithm.

\begin{figure}
\begin{center}
\includegraphics[width=0.8\textwidth]{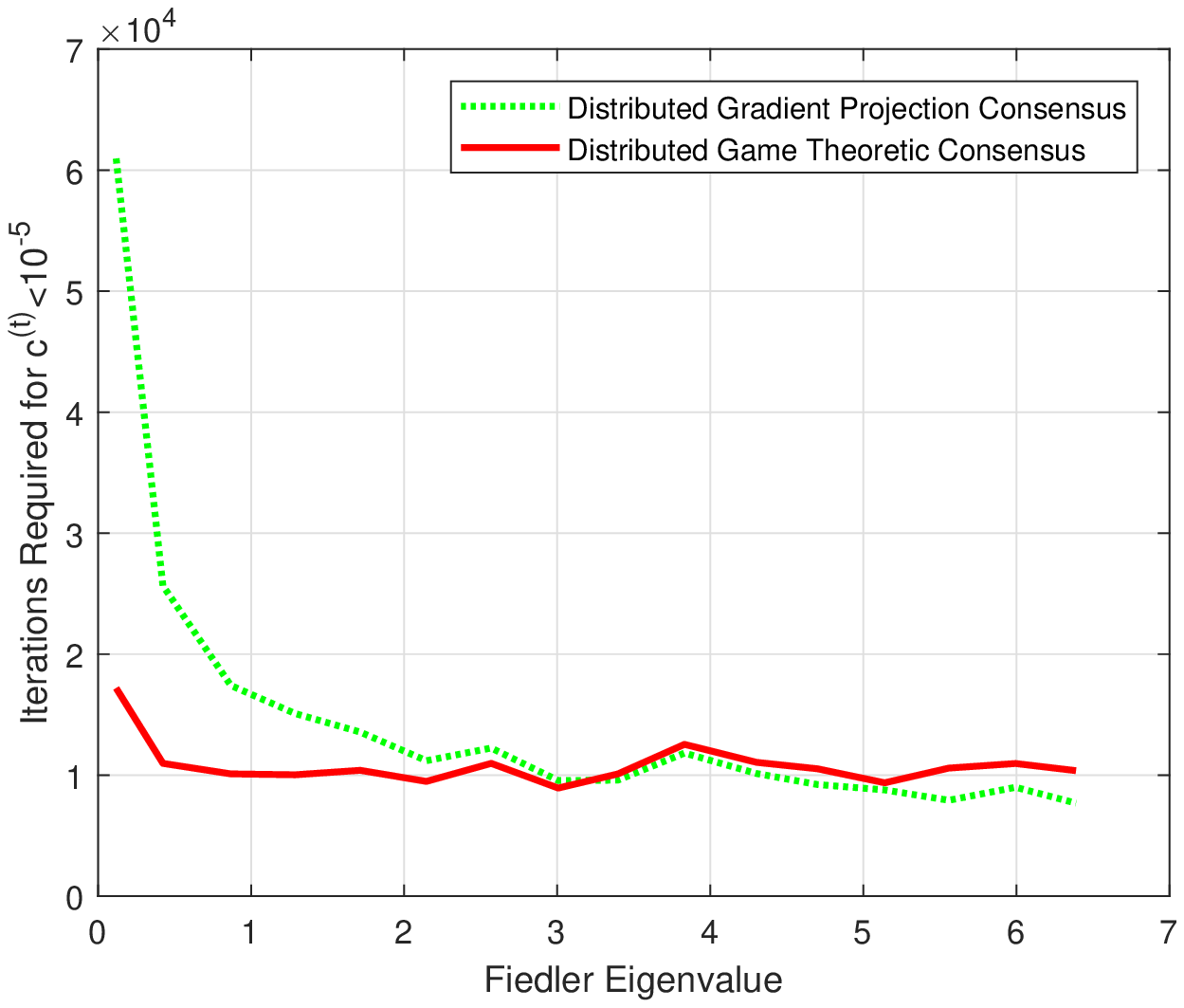}
\end{center}
\caption{Convergence rate comparison in $q=2$ dimensions for networks of various densities, communication range was uniformly selected in the interval $[0.1, 0.4]$}
\label{fig3}
\end{figure}

\begin{figure}
\begin{center}
\includegraphics[width=0.8\textwidth]{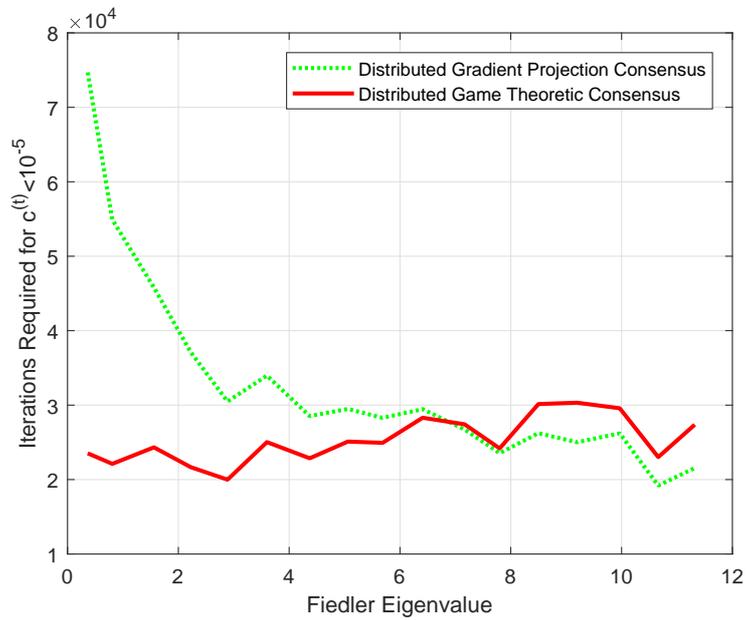}
\end{center}
\caption{Convergence rate comparison in $q=4$ dimensions for networks of various densities, communication range was uniformly selected in the interval $[0.4, 0.7]$}
\label{fig4}
\end{figure}

Fig. \ref{fig3} corresponds to a two dimensional ($q=2$) setting, where the communication range $\rho$ was uniformly varied in the interval $[0.1, 0.4]$. Similarly, Fig. \ref{fig4} corresponds to a four dimensional setting, where the communication range was uniformly varied in the interval $[0.4, 0.7]$. These figures demonstrate that the distributed game theoretic consensus algorithm (Table \ref{alg1}) converges faster than the distributed gradient projection consensus algorithm (Table \ref{alg2}), when the network topology is sparse enough, i.e. when the Fiedler eigenvalue of the network is smaller than 3 in the two dimensional case and when the Fiedler eigenvalue of the network is smaller than 7 in the four dimensional case. In more densely connected networks, the distributed gradient projection consensus algorithm that utilizes the step size given by equation (\ref{sim_step}) was found to converge faster.

\section{Conclusions}\label{sec:c}
In this work, the problem of computing a common point that lies in the intersection of a finite number of closed convex sets, each known to one agent in a network, was studied by following a game-theoretic approach. The problem was formulated as a non-cooperative game for which a potential function exists and it was proven that all Nash equilibria of this game correspond to consensus states. Following this result, a best-response based algorithm for distributed constrained consensus was derived. Furthermore, a projected gradient type algorithm was also considered. Finally, simulation results were given to validate the convergence of the studied schemes and to explore the relative convergence rates. The novel game-theoretic algorithm was shown to be particularly useful for consensus problems in sparsely connected networks.

\bibliographystyle{ieeetr}

\bibliography{ms}
	
\end{document}